\documentstyle[12pt]{article}
\begin{document} 
\vspace*{-1in} 
\renewcommand{\thefootnote}{\fnsymbol{footnote}} 
\begin{flushright} 
TIFR/TH/99-47\\
hep-ph/9909377\\
September 1999\\ 
\end{flushright} 
\vskip 65pt 
\begin{center} 
{\Large \bf Seeking TeV-Scale Quantum Gravity at an $e \gamma$ collider}\\
\vspace{8mm} 
{\bf 
Dilip~Kumar~Ghosh\footnote{dghosh@theory.tifr.res.in}, 
P.~Poulose \footnote{poulose@theory.tifr.res.in}, 
K.~Sridhar\footnote{sridhar@theory.tifr.res.in}
}\\ 
\vspace{10pt} 
{\sf Department of Theoretical Physics, Tata Institute of 
Fundamental Research,\\  
Homi Bhabha Road, Bombay 400 005, India. } 
 
\vspace{80pt} 
{\bf ABSTRACT} 
\end{center} 
Theories with large extra compact dimensions predict exciting phenomenological
consequences at the TeV scale. Such theories can, consequently, be 
tested/verified in experiments at future colliders like the Next Linear
Collider (NLC). In this paper, we study the production of the spin-2
Kaluza-Klein excitations in NLC operating in the $e \gamma$ mode. 
Results for unpolarised and polarised cases are presented and it
is shown that it is possible to use this process to test these theories
to values of the effective string scale between about 1 - 7 TeV.
\vskip12pt 
\noindent 
\setcounter{footnote}{0} 
\renewcommand{\thefootnote}{\arabic{footnote}} 
 
\vfill 
\clearpage 
\setcounter{page}{1} 
\pagestyle{plain}
\noindent In theories with large compact extra dimensions \cite{dimo,dimo2,shiu}, 
which have attracted much interest recently, the effects of gravity could 
become large at very low scales ($\sim$~TeV) and is of tremendous interest
to phenomenology. In these theories, $n$ of the dimensions of a higher 
dimensional string theory are compactified to a common scale $R$ which is large.
It is possible to have $R$ to be large because, in these theories,
only the gravitons (corresponding to closed strings) propagate in the 
$(4+n)$-dimensional bulk while the gauge particles (corresponding to open 
strings) live on a 3-brane and do not see the effects of the large extra
dimensions. The size of these extra dimensions are then restricted only
by gravitation experiments the constraints from which are relatively weak
\cite{gravexp} and allow the extra dimension to be as large as 1 mm.
The low-energy scale $M_S$ is related to the Planck scale by \cite{dimo}
\begin{equation} 
M^2_{\rm P}=M_{S}^{n+2} R^n ~,
\label{e1} 
\end{equation} 
It then follows that $R=10^{32/n -19}$~m, and so we find that $M_S$ can 
be arranged to be a TeV for any value $n > 1$.  It is at this low
scale of 1 TeV that we will now expect to see the effects of quantum
gravity~-- for example, for $n=2$ the compactified dimensions
are of the order of 1 mm, just below the experimentally tested region
for the validity of classical gravity and within the possible
reach of ongoing experiments. This very novel idea has far-reaching
consequences~: it is a possible solution to the heirarchy problem
(though the latter manifests itself in a new garb). But, more
interestingly, it is possible to make a viable scenario 
\cite{dimo4} which can survive the existing astrophysical and 
cosmological constraints and predict other interesting consequences like
low-energy unification \cite{dienes}.
For some early papers on large Kaluza-Klein dimensions, see Ref.~\cite{anto, 
taylor} and for recent investigations on different aspects of the
TeV scale quantum gravity scenario and related ideas, see Ref.~\cite{related}.

The low-energy effective theory that emerges below the scale $M_S$
\cite{sundrum, grw, hlz}, has an infinite tower of massive Kaluza-Klein
states, which contain spin-2, spin-1 and spin-0 excitations. For low-energy 
phenomenology the most important of these are the spin-2 Kaluza-Klein states 
i.e. the infinite tower of massive graviton states. The couplings of
the gravitons, $G_{\mu\nu}$, to the Standard Model (SM) fields 
are derivable from the following Lagrangian \cite{grw,hlz} :
\begin{equation} 
{\cal L}=-{1 \over \bar M_P} G_{\mu \nu}^{(j)}T^{\mu\nu} ~,
\label{e2} 
\end{equation} 
where $j$ labels the Kaluza-Klein mode, $\bar M_P=M_P/\sqrt{8\pi}$
and $T^{\mu\nu}$ is the energy-momentum tensor. Due to the sum over
the tower of graviton states, the Planck-scale suppression 
due to the $1/\bar M_P$ factor in the coupling is effectively replaced by
a TeV-scale ($\sim M_S$) suppression. Consequently, these couplings
can lead to observable effects at present and future colliders.
There have been several studies exploring these consequences.
Missing energy or monojet + missing energy signatures \cite{mpp, keung}
of real graviton production at $e^+ e^-$ or hadron colliders have been
studied which have yielded bounds on $M_S$ which are around 500 GeV to 
1.2 TeV at LEP2 \cite{mpp, keung} and around 600 GeV to 750 GeV at Tevatron 
(for $n$ between 2 and 6) \cite{mpp}. These studies have been extended to
the Large Hadron Collider (LHC) and to high-energy $e^+ e^-$ collisions
at the Next Linear Collider (NLC). Other studies have concentrated on the 
effects of virtual graviton exchange on experimental observables.
Virtual effects in dilepton production at Tevatron yields a bound of around 
950 GeV \cite{hewett} to 1100 GeV \cite{gmr} on $M_S$, in $t \bar t$ 
production at Tevatron a bound of about 650 GeV is obtained while at the 
LHC this process can be used to explore a range of $M_S$ values upto 
4~TeV \cite{us}. Virtual effects in deep-inelastic scattering at HERA 
put a bound of 550 GeV on $M_S$\cite{us2}, while from jet production at the 
Tevatron strong bounds of about 1.2 TeV  are obtained \cite{us3}.
More recently, fermion pair production and gauge boson production in $e^+ e^-$ 
collisions at LEP2 and NLC and in $\gamma \gamma$ collisions at the NLC 
[21 - 26]
have been studied. Effect of graviton
mediation Compton scattering at $e\gamma$ collider is considered in
\cite{davoudiasl2}. Associated 
production of gravitons with gauge bosons and virtual effects in gauge boson 
pair production at hadron colliders have also been studied \cite{balazs}.
Diphoton signals and global lepton-quark neutral current constraints have also
been studied \cite{cheung}. There have also been papers 
discussing the implications of the large dimensions for higgs
production \cite{rizzo2, xhe} and electroweak precision observables 
\cite{precision}. Astrophysical constraints, like bounds from energy loss 
for supernovae cores, have also been discussed \cite{astro}.

The Next Linear Collider (NLC) is an ideal testing ground of the SM and 
a very effective probe of possible physics that may lie beyond the SM. 
The collider is planned to be operated in the $e^+ e^-$, $\gamma \gamma$
and the $e \gamma$ modes. The photons are produced in the Compton 
back-scattering of a highly monochromatic low-energy laser beam off a 
high energy electron beam \cite{nlc}. Control over the $e^-$ and laser 
beam parameters allow for control over the parameters of the $\gamma \gamma$ 
and $e \gamma$ collisions. The physics potential of the NLC is manifold and 
the collider is expected to span several steps of $e^+ e^-$ energy between 
500 GeV and 1.5 TeV. The experiments at the NLC also provide a great degree 
of precision because of the relatively clean initial state, and indeed the 
degree of precision can be enhanced by using polarised initial beams. 

In the present paper, the effects of large extra dimensions are probed 
by looking for the production of the tower of spin-2 Kaluza Klein excitations 
in $e \gamma$ collisions at the NLC. The basic process that we are considering
is analogous to Compton scattering, except that the gauge boson in the final 
state is replaced by the graviton i.e. we have $e \gamma \rightarrow e G^m$,
where $G^m$ denotes a particular spin-2 state in the tower of excitations.
The $e \gamma$ scattering can be thought of as a subprocess of the
primary scattering of the $e^-$ and the $e^+$, with 
the $\gamma$ being produced from the $e^-$ (or $e^+$)-laser back
scattering. The energy of the back-scattered photon, $E_\gamma$, follows
a distribution characteristic of the Compton scattering process and
can be written in terms of the dimensionless ratio $x=E_\gamma/E_e$.
It turns out that the maximum value of $x$ is about 0.82\footnote{
Above this value the scattered photon interaction with the laser photon
start producing $e^+e^-$ pairs, which in turn reduces luminosity.}
so that provides
the upper limit on the energy accessible in the $e \gamma$ sub-process.
We assume that the primary $e^+ e^-$ scattering process can be factorised
into a part that describes the $e \gamma$ subprocess scattering and 
a luminosity function, $f_\gamma (x)$, where the latter provides 
information on the photon flux produced in Compton scattering of the 
electron and laser beams \cite{lumino}. 

We calculate the cross-section for the process $e \gamma \rightarrow e G^m$,
where $G^m$ is a definite spin-2 state of mass $m$. In order to obtain the 
cross-section for the entire range of accessible states we will have to 
sum this cross-section over these states. For small enough mass-splittings, 
this sum can be replaced by an integral over the mass parameter $m$. 
The resulting inclusive graviton production cross-section can be written
in terms of the cross-section for the process $e \gamma \rightarrow e G^m$ 
as follows:
\begin{equation}
\frac{d^2\hat{\sigma}}{dm\,dt}=S_{n-1}
	\frac{m^{n-1}}{M_S^{2+n}}\,\overline{M}_P^2\,
	\frac{d\hat{\sigma}^{m}}{dt}  ,  \\ 
\label{e3} 
\end{equation}
where $n$ is number of extra dimensions, and for $n=2k$ where $k$ is an 
integer, $S_{n-1}=2\pi^k/(k-1)!$ and for $n=2k+1$, $S_{n-1}=2\pi^k/
\Pi^{k-1}_{j=0} (j+\frac{1}{2})$. $d\hat{\sigma}^{m}/d\hat{t}$
is the differential cross-section for producing a single graviton state of
mass $m$ and is given by 
\begin{equation}
\frac{d\hat{\sigma}^{m}}{dt}=\frac{1}{16 \pi s^2}\left[
	\frac{1}{2}(1+P_e\xi_2)\,|{\cal M}(++)|^2+
	\frac{1}{2}(1-P_e\xi_2)\,|{\cal M}(+-)|^2 \right] ,
\label{e4} 
\end{equation}
where $P_e$ is the rate of polarisation of the $e$ and $\xi_2$ is the Stokes 
parameter which defines the polarisation state of the photon and is
fixed in terms of the polarisations of the initial electron and laser
beam. The helicity amplitudes are given by 
\begin{eqnarray}
|{\cal M}(++)|^2 &=&\frac{2\pi \alpha}{\overline{M}_P^2}\frac{1}
	{s t u}(m^2-s)^2 (4 su - m^2t ) \nonumber \\
|{\cal M}(+-)|^2 &=&\frac{2\pi \alpha}{\overline{M}_P^2}\frac{1}
	{s t u} (s + t)^2 (4 s u - m^2 t).
\label{e5} 
\end{eqnarray}
From Eqns.~\ref{e3}, \ref{e4}, \ref{e5}, one sees explicitly that 
the factor $\overline{M}_P^2$ from the phase space summation compensates 
for the $\overline{M}_P$ dependence of the cross-section. One can then
use the inclusive cross-section in Eq.~\ref{e3} to write down the
differential cross-section in transverse momentum $p_T$ and rapidity $y$
of the final-state electron as a convolution over the photon luminosity
function:
\begin{equation}
\frac{d^2\sigma}{dp_T^2dy}=\int dx\,f(x)\,\frac{s}{2m}\,
	\left(\frac{d^2\hat{\sigma}}{dm\,dt}\right) \,
	\delta(s+t+u-m^2)\,dm .\\ 
\label{e6} 
\end{equation}

If the graviton is produced in the $e \gamma$ collision it will escape
detection giving rise to a missing energy signature with an isolated
electron in the final state. The SM background to this signal comes
from $e\gamma \rightarrow e Z$ followed by $Z \rightarrow \nu \bar\nu$
and $e\gamma \rightarrow \nu W$ with $W \rightarrow e \nu$. This SM
background has been studied extensively in the context of selectron/
neutralino production at $e \gamma$ collider \cite{backgd}. We have
analysed the signal and the SM background for $e^+ e^-$ centre-of-mass
energies corresponding to 500, 1000 and 1500 GeV, respectively. For
our numerical results, we have assumed an integrated luminosity, $\cal{L}$, of 
100 fb${}^{-1}$.

\begin{table}
\begin{center}
\begin{tabular}{|c|c|c|c|c|c|}
\hline
&\multicolumn{5}{|c|}{ }\\
$\sqrt{s}_{ee}$&\multicolumn{5}{|c|}{$M_S$ in GeV}\\[3mm]
\cline{2-6}
&&&&&\\
TeV&$n=2$&3&4&5&6\\[3mm]
\hline
&&&&&\\
0.5&2267&1648&1316&1111&971\\
1.0&3192&2508&2109&1847&1658\\
1.5&3907&3205&2776&2479&2260\\[3mm]
\hline
\end{tabular}
\caption{\it Limit on $M_S$ in the case of unpolarized electron and photon 
beams for different c.m.f energies ($\sqrt{s}_{ee}$). $n$ is the 
number of compactified extra dimensions. A geometric luminosity of 100 
fb$^{-1}$ is assumed and a $p_T$ cut of 10 GeV and a rapidity cut above 
3 are applied. }
\end{center}
\end{table}

For our analysis, we have used a cut of 10 GeV on the electron $p_T$.
We demand a statistical significance of the signal defined as 
$\sigma \sqrt{\cal L} /\sqrt{B}$ to be greater than 5, where $\sigma$ 
denotes the signal cross-section and $B$ the SM background cross-section.
Using this criterion we get the bounds on $M_S$, the results for the
unpolarised case are given in Table 1, where the reach in $M_S$ is shown
for different values of the centre-of-mass energy and the number of extra
dimensions. In the unpolarised case, the biggest problem is due to the 
$W$ background, which is substantially larger than the signal as well 
as the $Z$ background. One way of getting rid of the $W$ contamination
is to use a right-handed electron. With a right-polarised electron (but
with photon still unpolarised) there is a substantial improvement in
the reach in $M_S$ -- an improvement which is as high as 60\% for some
values of $\sqrt{s}_{ee}$ and $n$. The corresponding results are 
tabulated in Table 2. 

\begin{table}
\begin{center}
\begin{tabular}{|c|c|c|c|c|c|}
\hline
&\multicolumn{5}{|c|}{ }\\
$\sqrt{s}_{ee}$&\multicolumn{5}{|c|}{$M_S$ in GeV}\\[3mm]
\cline{2-6}
&&&&&\\
TeV&$n=2$&3&4&5&6\\[3mm]
\hline
&&&&&\\
0.5&2939&2028&1565&1289&1105\\
1.0&4713&3425&2735&2307&2015\\
1.5&6245&4665&3793&3241&2858\\[3mm]
\hline
\end{tabular}
\caption{\it Limit on $M_S$ in the case of right-polarized electron beam and
unpolarized photon beam.  All other parameters are kept the same ase in 
the case of Table 1.}

\end{center}
\end{table}

In Fig.~1, we have plotted the rapidity distribution of the electron
for the case of the signal (full line) and the $Z$ background (dashed 
line). We find that the $Z$ background is peaked in the negative rapidity
region, which suggests that it will be expedient to use a positive $y$
cut so as to diminish the effects of the $Z$ background. With a positive
$y$ cut, we find again a moderate enhancement in the values of $M_S$
that can be bounded.

\begin{figure}[ht]
\vspace*{3.7in}
\includegraphics{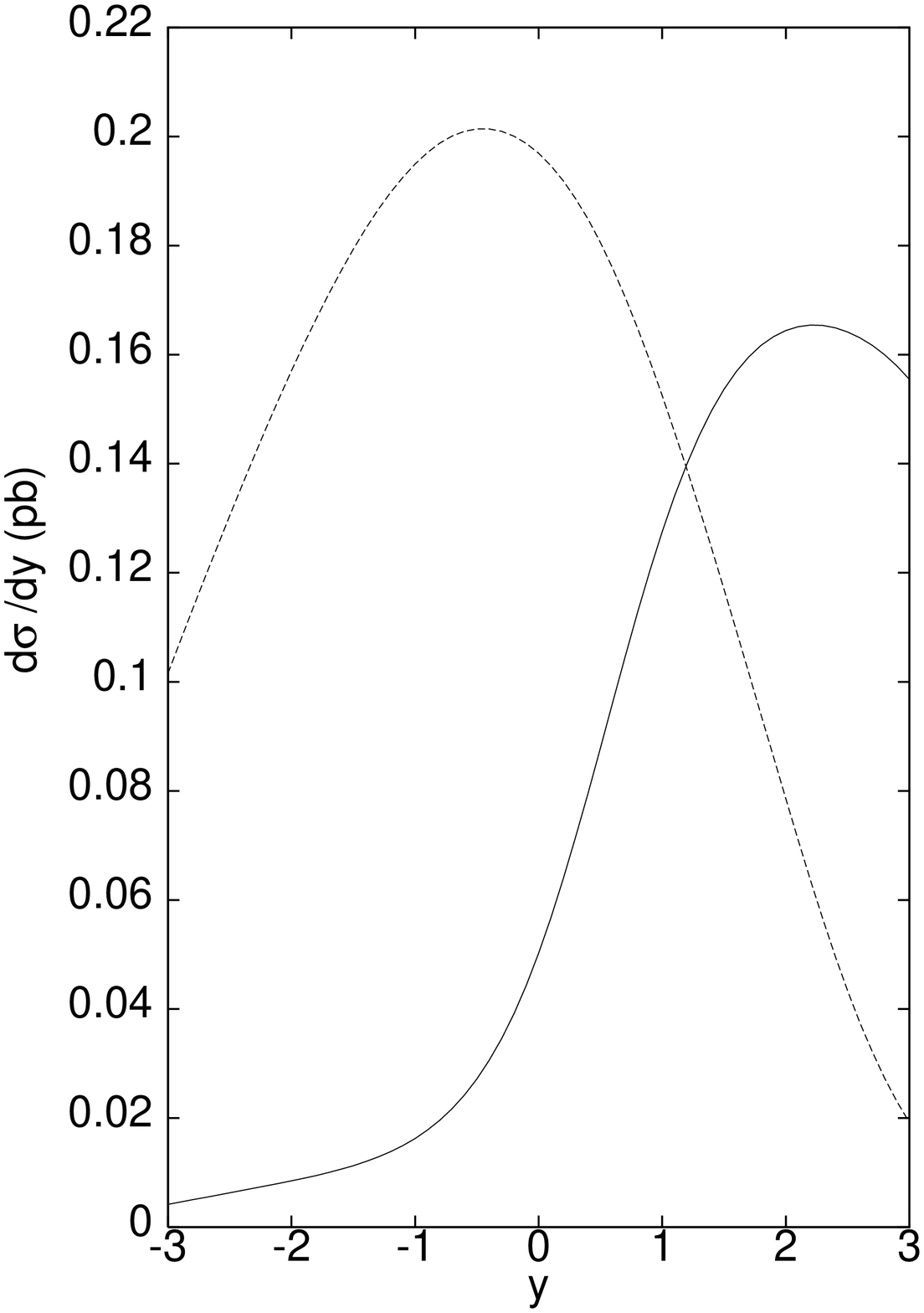}
\caption{\it The electron $y$ distribution for
$\sqrt{s}_{ee}=1000$
GeV and for a right-handed initial electron and unpolarized photon. 
The solid line is the signal
due to the graviton production and the dashed line is the SM background
from Z production. Signal is for an $M_S$ value of 1 TeV. A $p_T$ cut of
10 GeV is assumed.} 
\end{figure}

Finally, we present the results for the polarised case, where the electron 
is polarised and the $\gamma$ is also polarised. 
For the polarised case, we study the singal and the SM background 
for different choices of the initial electron and laser beams. For a
given choice of the $e^-$ and laser polarisation, the photon polarisation 
is fixed once the $x$ value is known. The latter polarisation is therefore
dependent crucially on the luminosity functions and it is only on the
polarisation of the electron and the laser beams that we have a direct
handle. The efficacy of polarisation as a discriminator of the new physics
is, however, apparent more at the level of the $e \gamma$ sub-process.
As we scan over the different choices of initial beam polarisations, we
find that for certain choices there is hardly any sensitivity to the
new physics. However, large differences are realised for certain other
choices. After scanning over the possible values of the photon polarisation,
we find that for a particular choice of the initial electron and laser
polarisations the best sensitivity results. We present our results for
this particular case in Table 3. Again we find that for this choice 
of polarisation a much stronger bound on $M_S$ results as compared
to the unpolarised case.

\begin{table}
\begin{center}
\begin{tabular}{|c|c|c|c|c|c|}
\hline
&\multicolumn{5}{|c|}{ }\\
$\sqrt{s}_{ee}$&\multicolumn{5}{|c|}{$M_S$ in GeV}\\[3mm]
\cline{2-6}
&&&&&\\
TeV&$n=2$&3&4&5&6\\[3mm]
\hline
&&&&&\\
0.5&3369&2274&1721&1394&1180\\
1.0&5712&4018&3122&2576&2211\\
1.5&7769&5587&4406&3674&3177\\[3mm]
\hline
\end{tabular}
\caption{\it Limits on $M_S$ in the case of righ-polarized electron beam and
photon beam obtained by the Compton back scattering of left-handed
electron and left-circularly polarized laser photon. A geometric
luminosity of 100 fb$^{-1}$ is assumed and a $p_T$ cut of 10 GeV and
rapidity ($y$) cut of 3 are applied.}
\end{center}
\end{table}

In conclusion, the physics of large extra dimensions can be tested very
effectively at the Next Linear Collider. In particular, in the present
paper, we have considered the production of spin-2 Kaluza-Klein excitations
in $e \gamma$ collisions. We find that in the unpolarised case
$M_S$ values upto a few TeV are probed in this process. The use of 
polarisation helps to strengthen these bounds quite significantly.
We have shown that it is possible to reduce the SM backgrounds by
choosing a right-polarised electron or by choosing a suitable rapidity
cut. A more detailed study than the one presented here can be used
to tighten the kinematical cuts  and improve the discriminatory power
of this process.

\end{document}